\documentclass[aps,prx, reprint, superscriptaddress, footinbib]{revtex4-1}
\usepackage[latin1]{inputenc}
\usepackage{graphicx}
\usepackage{bm}
\usepackage{color}
\usepackage{verbatim}
\usepackage{placeins}
\usepackage{soul}
\usepackage{amsmath}

\newcommand{\kB}{\ensuremath{k_\mathrm{B}} }

\newcommand{\sigmaz}{\ensuremath{\sigma_{\mathrm{\mathrm{z}}}}}
\newcommand{\sigmax}{\ensuremath{\sigma_{\mathrm{\mathrm{x}}}}}

\newcommand{\Hq}{\ensuremath{H_\mathrm{q}}}
\newcommand{\omegaq}{\ensuremath{\omega_{\mathrm{\mathrm{q}}}}}

\newcommand{\EJ}{\ensuremath{E_\mathrm{J}}}
\newcommand{\EC}{\ensuremath{E_\mathrm{C}}}
\newcommand{\Ic}{\ensuremath{I_\mathrm{c}}}
\newcommand{\nng}{\ensuremath{n_\mathrm{g}}}

\newcommand{\omegam}{\ensuremath{\Omega}}			
\newcommand{\Hm}{\ensuremath{H_\mathrm{m}}}
\newcommand{\prob}{\ensuremath{p_+}}
\newcommand{\rhoM}{\ensuremath{\rho_\mathrm{m}}}
\newcommand{\Qm}{\ensuremath{Q_\mathrm{m} }}
\newcommand{\XZP}{\ensuremath{X_\mathrm{ZP}}}

\newcommand{\Trm}{\ensuremath{\mathrm{Tr}_\mathrm{m}}}

\newcommand{\ad}{\ensuremath{a^\dagger}}
\newcommand{\bra}[1]{\ensuremath{\left\langle #1 \right|}}

\newcommand{\ket}[1]{\ensuremath{\left| #1 \right\rangle}}

\newcommand{\resonator}{resonator }
\newcommand{\resonators}{resonators }

\begin{document}
	
\title{Displacemon electromechanics: how to detect quantum interference in a nanomechanical \resonator}

\author{K.~E.~Khosla}
\affiliation{Center for Engineered Quantum Systems, The University of Queensland, Brisbane, Queensland 4067, Australia}
\affiliation{School of Mathematics and Physics, The University of Queensland, Brisbane, Queensland 4067, Australia}

\author{M.~R.~Vanner}
\affiliation{QOLS, Blackett Laboratory, Imperial College London, London SW7 2BW}
\affiliation{Clarendon Laboratory, Department of Physics, University of Oxford, Oxford OX1 3PU, United Kingdom}

\author{N.~Ares}
\affiliation{Department of Materials, University of Oxford, Parks Road, Oxford OX1 3PH, United Kingdom}

\author{E.~A.~Laird}
\email{edward.laird@materials.ox.ac.uk}
\affiliation{Department of Materials, University of Oxford, Parks Road, Oxford OX1 3PH, United Kingdom}

\begin{abstract}
We introduce the `displacemon' electromechanical architecture that comprises a vibrating nanobeam, e.g. a carbon nanotube, flux coupled to a superconducting qubit. This platform can achieve strong and even ultrastrong coupling enabling a variety of quantum protocols. We use this system to describe a protocol for generating and measuring quantum interference between two trajectories of a nanomechanical resonator. The scheme uses a sequence of qubit manipulations and measurements to cool the resonator, apply an effective diffraction grating, and measure the resulting interference pattern. We simulate the protocol for a realistic system consisting of a vibrating carbon nanotube acting as a junction in a superconducting qubit, and we demonstrate the feasibility of generating a spatially distinct quantum superposition state of motion containing more than $10^6$ nucleons. 
\end{abstract}

\date{\today{}}
\maketitle

\section{Introduction}

The superposition principle is a fundamental tenet of quantum mechanics and essential for understanding a wide range of quantum phenomena. As the scale of quantum objects increases, the experimental consequences of this principle become increasingly hard to isolate.  Is there a scale at which this tenet begins to break down? The strongest tests of superposition come from matter-wave interferometry between trajectories of large molecules. Remarkably, interference can be measured using molecules of mass as large as $7\times 10^3$~atomic mass units (amu)~\cite{Hornberger2012,Eibenberger2013}. The ability to create unambiguous superpositions on a mesoscopic scale would allow  tests of quantum collapse theories~\cite{Nimmrichter2013} and gravitational decoherence~\cite{Diosi1984,Penrose1996}, ultimately addressing experimentally the question of why we fail to see superpositions in everyday life~\cite{Bassi2013}. This has inspired numerous challenging proposals to detect interference of larger particles~\cite{Marshall2003,Romero-Isart2011,Bateman2013} via optomechanical coupling~\cite{Aspelmeyer2014}, or using levitated nanodiamonds~\cite{Yin2013,Scala2013}.

Nanomechanical \resonators span this mesoscopic mass scale from $\sim 10^6$ to $\sim10^{16}$ amu and therefore provide an attractive route to extend the scale over which quantum effects can be observed.
Recently, cooling to the ground state~\cite{OConnell2010,Chan2011,Teufel2011}, and
such elements of quantum behaviour as state squeezing~\cite{Wollman2015,Lecocq2015,Pirkkalainen2015} and coherent qubit coupling~\cite{OConnell2010,Pirkkalainen2013,Lecocq2014} have become accessible with mechanical \resonators of this scale. 
Moreover, significant progress towards observing mechanical superposition states has been made in both opto- and electromechanics, and mechanical interference fringes have recently been observed at a classical level~\cite{Ringbauer2016}. The observation of quantum interference, however, remains outstanding and is a key goal of this paper.

Here we introduce the	``displacemon'', a device that enables strong coupling between a nanomechanical \resonator and a superconducting qubit. We show how to create an effective diffraction grating that leads to an interference pattern in the resonator's displacement. The scheme works using a sequence of manipulations on the qubit to create an effective grating with a fine pitch and therefore a large momentum displacement. In molecular interference experiments, the diffraction grating is typically an etched membrane; however, van der Waals interactions with the slits mean this is hard to extend to large particles~\cite{Hornberger2012}. More advanced implementations use optically defined gratings; the pitch, which sets the momentum separation of the diffracted beams, is then limited by the optical wavelength~\cite{Arndt2014}. In our scheme, the pitch is limited neither by an optical wavelength nor by the size of the \resonator, but by the qubit-\resonator coupling strength. As we will show, this allows for diffraction gratings with a pitch narrower than the ground state wave function.

Our proposed device uses a vibrating nanobeam flux-coupled to a superconducting qubit, through which all manipulations and measurements are performed. As a nanobeam that optimally combines high mechanical frequency, low dissipation, and the ability to couple strongly to superconducting quantum devices, we propose a suspended carbon nanotube. Previous proposals for quantum motion in nanotubes~\cite{Palyi2012, Ohm2012} have been based on coupling to a spin qubit; however, the coherence requirement on the qubit is stringent~\cite{Pei2017}. Here, using realistic parameters derived from experiments, we show how to construct an effective mechanical diffraction grating and measure quantum interference in a moving object of $>10^6$~amu. This work enables the mass scale on which quantum interference can be observed to be extended by nearly three orders of magnitude. 

\section{Model}

\begin{figure}
	\includegraphics[width=86mm]{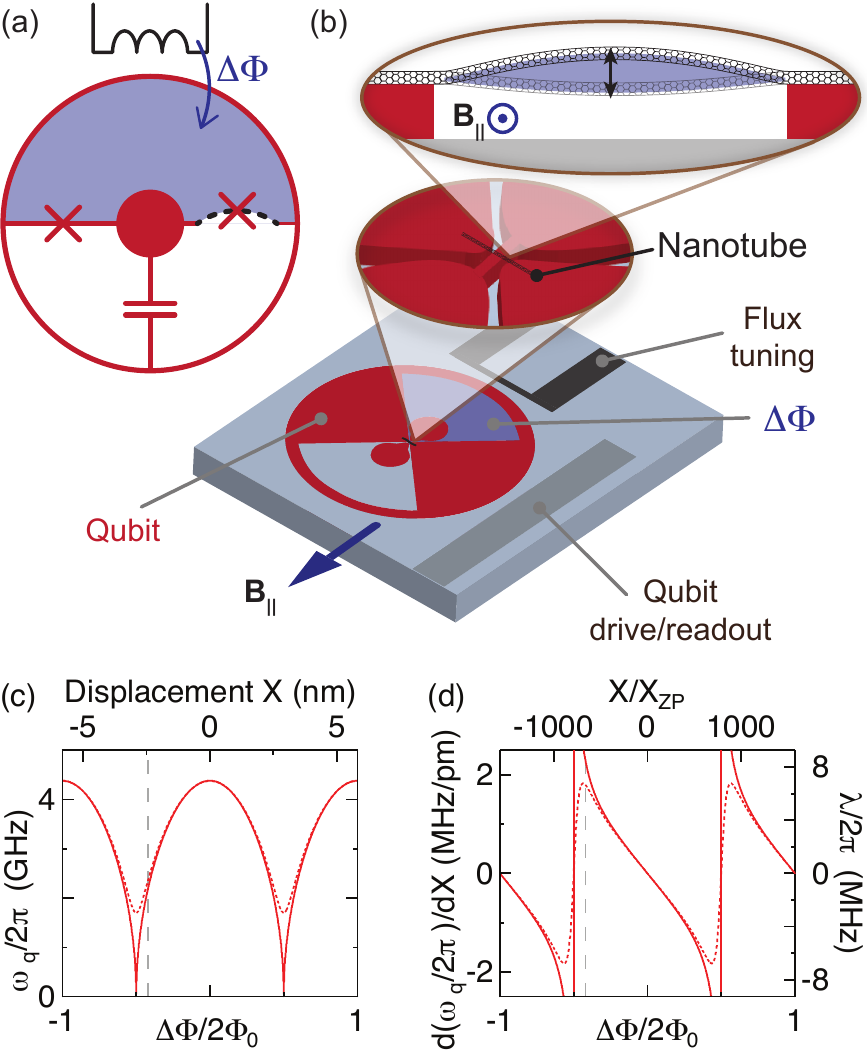}
	\caption{\label{Fig1} Device for strong qubit-mechanical coupling. (a) Electrical schematic. The device is a gradiometric transmon qubit, biased by flux difference $\Delta \Phi$ between the SQUID loops. With a suspended nanotube acting as at least one junction (shown here for the right junction), the displacement modulates $\Delta \Phi$ and therefore the qubit levels. (b) Arrangement of the qubit, vibrating nanotube, flux tuning coil, and drive/readout cavity antenna. The in-plane magnetic field $B_{||}$ introduces strong coupling between the vibrations and $\Delta \Phi$.  (c) Qubit frequency as a function of flux difference, with parameters as in the text. Solid lines assume equal Josephson coupling in the two SQUID junctions, dotted line assumes 30~\% asymmetry (see Appendix~\ref{app:parameters}). Curves are plotted as a function of flux (bottom axis) and equivalently of displacement (top axis). (d) Qubit displacement sensitivity (left axis) and mechanical coupling rate (right axis) as a function of flux. The bias point that achieves the assumed coupling is indicated by a vertical dashed line.}
\end{figure}

In general, strongly coupling a mechanical \resonator to a qubit is challenging because the best qubits are engineered to be insensitive to their environment~\cite{Treutlein2014}. 
We propose a design that is robust against electrical and magnetic noise, while still achieving strong mechanical coupling. We envisage a superconducting qubit of the concentric transmon design~\cite{Braumuller2016} in which at least one of the junctions is a vibrating nanotube (Fig.~\ref{Fig1}). Nanotube \resonators offer unique advantages for studying quantum motion~\cite{Sazonova2004}: (i) the zero-point amplitude is typically greater than 1~pm, much larger than other mechanical \resonators; (ii) the resonant frequency is sufficiently large to allow near-ground-state thermal occupation, suppressing thermal decoherence~\cite{Laird2012, Huttel2009}; (iii) a nanotube can act as a Josephson junction~\cite{Jarillo-Herrero2006,Cleuziou2006,Schneider2012}; and (iv) ultraclean devices offer quality factors greater than $10^6$, which provide long-lived mechanical states~\cite{Moser2014}.

In this design (Fig.~\ref{Fig1}(a)), the two junctions form a gradiometric superconducting quantum interference device (SQUID), so that the qubit frequency is set by the flux difference $\Delta \Phi$ between the two loops. This flux difference is tuned primarily by means of a variable perpendicular field $\Delta B_\bot(t)$, while  mechanical coupling is flux-mediated using a static in-plane field $B_{||}$~\cite{Xue2007,Etaki2008,Shevchuk2017}. This type of concentric transmon is insensitive to uniform magnetic fields, which has two advantages for this proposal: the qubit can be operated coherently away from a flux sweet spot~\cite{Braumuller2016}, and any misaligned static flux does not perturb the energy levels. Both these facts are favourable for strong nanomechanical coupling. Because this variant of the transmon is designed for strong coupling to nanomechanical displacement, we refer to it as a `displacemon'.
In this section we derive the displacemon Hamiltonian, and estimate the parameters for a feasible device.

\subsection{The mechanical \resonator}
We consider the nanotube as a beam of length $l$ and diameter $D$ and focus our studies on its fundamental vibrational mode~\cite{Sazonova2004, Ustunel2005, Witkamp2006, Huttel2009, Laird2012, Moser2014, Ares2016}. The mechanical resonator Hamiltonian is
$
\Hm = \hbar \Omega \ad a,
$
where $\ad\,(a)$ is the creation (annihilation) operator for the \resonator.
Typically, the restoring force for a clamped nanotube is dominated by the beam's tension~$T$~\cite{Poot2012}, so that the mechanical angular frequency is $\omegam = \frac{\pi}{l} \sqrt{\frac{T}{\mu}}$ and the zero-point amplitude is $\XZP = \sqrt{\frac{\hbar}{2m\omegam}} = \sqrt{\frac{\hbar}{2\pi}}\, (\mu T)^{-1/4}$, where $\mu = \pi\rho_\mathrm{S}D$ is the mass per unit length and $\rho_\mathrm{S} = 8 \times 10^{-7}~\mathrm{kg~m}^{-2}$ is the sheet density of graphene. The displacement profile as a function of axial coordinate $Z$ is
$
\tilde{X}(Z) = X\sqrt{2} \sin \frac{\pi Z}{l},
$
where $X \equiv (a + \ad)\XZP$ is the displacement coordinate.
This profile is normalized so that the root mean square displacement is equal to~$X$~\cite{Poot2012}. The flux coupling is proportional to the area swept out by the nanotube, which is equal to $\beta_0 l X$, where $\beta_0\equiv \frac{1}{l X}\int_0^l \tilde{X}(Z)\,dZ=\frac{2\sqrt{2}}{\pi}$ is a geometric coupling coefficient~\cite{Etaki2008}.

Nanotube \resonators can also be fabricated without tension, so that the restoring force is dominated by the beam's rigidity~\cite{Ustunel2005, Witkamp2006}. In this limit, the mechanical frequency is $\omegam=\frac{22.4}{l^2}\sqrt{ED^2/8\mu}$ and the coupling coefficient is $\beta_0 = 0.831$, where $E\approx D \times 1.09\times 10^{3}~\mathrm{Pa\,m}$ is the extensional rigidity.

\subsection{The qubit}
The qubit consists of a pair of superconducting electrodes coupled through the SQUID junctions. The qubit Hamiltonian is~\cite{Koch2007}
\begin{equation}
\Hq = 4\EC (\hat{n} - \nng)^2 - \EJ \cos \hat\varphi ,
\label{eq:transmonfullH}
\end{equation}
where $\EC$ is the charging energy, $\EJ$ is the SQUID Josephson energy, and $\hat n$ and $\hat\varphi$ are the overall charge (expressed in Cooper pairs) and phase across the junctions, with $\nng$ being the offset charge. Here we have neglected the qubit inductance, which makes a small contribution on the energy levels~\cite{Braumuller2016}.
In the transmon limit $\EJ \gg \EC$, we can approximate  Eq.~(\ref{eq:transmonfullH}) by an effective Hamiltonian
$
\Hq \approx \frac12 \hbar \omegaq \sigma_z ,
$
where $\omegaq = \sqrt{8\EJ \EC}/\hbar$ is the qubit frequency and $\sigma_z$ is the standard Pauli matrix, acting on the qubit ground state $|-\rangle$ and the excited state $|+\rangle$.  Qubit rotations, initialization, and projective measurement are now well-established through capacitive coupling to a microwave cavity in a circuit quantum electrodynamics architecture~\cite{Mallet2010,Braumuller2016}.

\subsection{Strong and ultrastrong coupling}
Strong and tunable coupling between the qubit and the mechanical \resonator is achieved by flux coupling to the SQUID loops, which tunes the qubit Josephson energy. Assuming equal critical current $\Ic$ in the two junctions, this Josephson energy is
\begin{equation}
\EJ = \EJ^0 \left|\cos \frac{\pi \Delta \Phi}{2\Phi_0}\right|,
\label{eq:EJ}
\end{equation}
where $\Delta \Phi$ is the flux difference between the two loops, $\EJ^0=\Ic\Phi_0/\pi$ is the maximum Josephson energy, and $\Phi_0=h/2e$ is the flux quantum.

The flux difference can be tuned both directly, via a perpendicular magnetic field $B_\bot$, and via the displacement using a static in-plane field $B_{||}$. We have 
\begin{equation}
\Delta \Phi = A \Delta B_\bot + 2 \beta_0 l B_{||} X,
\end{equation} where $A$ is the area of one SQUID loop. Since quite small perpendicular fields suffice to tune the qubit frequency over its full range, we envisage an on-chip coil to modulate $\Delta B_\bot(t)$ as a function of time $t$~\cite{DiCarlo2009,Braumuller2016}. 
Substituting Eq.~(\ref{eq:EJ}) into the definition of $\omegaq$ gives:
\begin{equation}
\frac{d \omegaq}{dX} = - \omegaq^0 \frac{\pi \beta_0 l B_{||}}{2\Phi_0} \frac{\sin \pi \Delta\Phi/2\Phi_0}{\sqrt {|\cos \pi \Delta\Phi/2\Phi_0|}},
\label{eq:domegadX}
\end{equation}
where $\omegaq^0 = \sqrt{8 \EJ^0 \EC}/\hbar$ is the maximal qubit frequency.
The dependence of $\omegaq$ on $X$ gives rise to an electromechanical coupling, resulting in the Hamiltonian~\cite{Armour2002,Treutlein2014}:
\begin{eqnarray}
H = \hbar\omegam \ad a + \hbar\frac{\omegaq}{2} \sigmaz +\hbar \frac{\lambda(t)}{2} (a + \ad) \sigmaz ,
\label{eq:Hamiltonian}
\end{eqnarray}
where $\lambda(t) = \XZP \, d\omegaq/dX$ (from Eq.~\eqref{eq:domegadX}) is the 
qubit-mechanical coupling strength, dynamically controlled through the field $\Delta B_\bot(t)$. 

To achieve coherent interaction between the qubit and the \resonator requires the strong coupling regime, where the maximum accessible coupling $\lambda_0$ exceeds the thermal decay rate $\kappa_{\mathrm{th}}=\kB T/(\hbar\Qm)$ of the \resonator, where $\Qm$ is the quality factor, and the decoherence rate of the qubit $\gamma=1/T_2$, where $T_2$ is the coherence time. The large zero-point motion makes nanotube \resonators particularly favourable for achieving this regime. Taking device parameters from simulation and experiment (Appendix~\ref{app:parameters}) leads to $\Omega/2\pi = 125$~MHz, $\omegaq/2\pi = 2.19$~GHz, and $\lambda_0/2\pi = 8.5$~MHz, with the flux dependence shown in Fig.~\ref{Fig1}. This is favourable for achieving the strong coupling regime, since both $\kappa_{\mathrm{th}}/2\pi$ and $\gamma/2\pi$ are typically less than 1~MHz.

To create well-separated mechanical superpositions, a stronger condition is desirable; the qubit should precess appreciably within an interval during which the \resonator can be considered stationary. This is the ultrastrong coupling regime, where $\lambda_0 > \omegam$~\cite{Shevchuk2017}. It is possible that a device similar to that of Fig.~\ref{Fig1} could access this regime (see Appendix~\ref{app:parameters}). However, here we instead suppose that effective ultrastrong coupling is engineered by modulating $\Delta B_\bot(t)$ at the mechanical frequency (see Section~\ref{sec:toggling}). In this modulated frame (similar to the toggled frame obtained by repeatedly flipping the qubit~\cite{Kolkowitz2012}), the \resonator is effectively frozen~ and the ultrastrong coupling condition is relaxed to $\lambda_0 > \kappa_{\mathrm{th}},\gamma$.

\section{Generating and measuring mechanical quantum interference}

To realise the nanomechanical interferometer, we propose a series of operations and measurements on the qubit. The qubit provides the necessary non-linearity to generate mechanical superposition states. The core idea is that the state of the \resonator is constrained by the qubit measurement outcome in the same way that the state of a particle is constrained by passing through a diffraction screen. By concatenating a series of qubit rotations and measurements, the \resonator can be cooled, diffracted and measured. 

\subsection{Cooling the \resonator}

\begin{figure}
	\includegraphics[width=86mm]{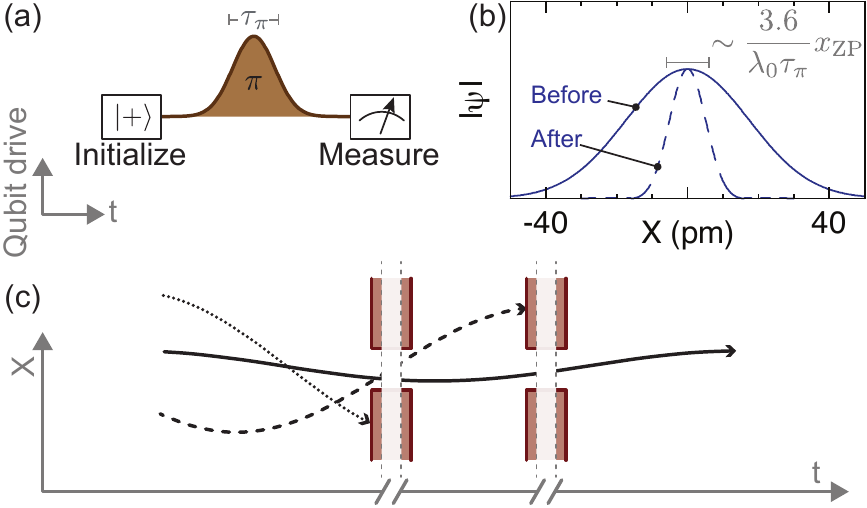}
	\caption{\label{Fig2} Cooling the \resonator using the qubit. (a) Pulse sequence (see text). Gaussian pulse shaping is chosen for near-optimal filtering. (b) Effect of this sequence, conditioned on the qubit outcome $|-\rangle$, on an initial state with a thermal distribution $\overline{n}=5$. (Burst duration is $\tau_\pi=100$~ns with $\lambda_0/2\pi = 800$~kHz.) The pulse sequence has the effect of passing the wavefunction through a narrow slit. The filter function is obtained by numerically solving the time evolution generated by Eq.~\eqref{eq:Hamiltonian} with an additional $\pi$-pulse term $g(t)\sigmax/2$. (c) Cartoon showing the effect of two pulse sequences, offset by a quarter-integer number of mechanical periods. Only the lowest-energy trajectories (solid line) survive both measurements, effectively cooling the \resonator. For the purpose of this cartoon, the duration of the two pulse sequences is compressed; in fact, each pulse sequence lasts for several mechanical periods.}
\end{figure}

The first step is to prepare the \resonator close to its ground state. A mechanical frequency of 125~MHz requires a bath temperature below $\sim 5$~mK for a thermal occupation less than unity. Such temperatures are achievable but challenging with cryogenic cooling~\cite{Bradley2016}. At a more accessible cryostat temperature of 33~mK, the initial thermal occupation is $\overline{n}=5$. To approach the ground state from a thermal state, we propose here an active cooling scheme utilising the qubit as a thermal filter~(Fig.~\ref{Fig2}). Following initialization to the $|+\rangle$ state, the scheme consists of applying a $\pi$ burst at the bare qubit frequency $\omegaq$~(Fig.~\ref{Fig2}(a)). If the \resonator is near its equilibrium position, this results in a qubit flip. By conditioning on this outcome (i.e., utilising only those runs of the experiment where this qubit outcome is measured), the \resonator state is constrained to a narrow window~(Fig.~\ref{Fig2}(b)).

A single operation of this type cools only one quadrature of the motion, because \resonator states with high momentum may still pass the window. To cool the orthogonal quadrature, the same selection should be applied a quarter of a mechanical period later~\cite{Vanner2011}, which filters out high-energy states that pass the first selection step~(Fig.~\ref{Fig2}(c)). The combination of these two pulse sequences can therefore prepare the \resonator close to its ground state, at the price of accepting only a fraction of the measurement runs.

\subsection{Diffracting the \resonator}

\begin{figure}
	\includegraphics[width=86mm]{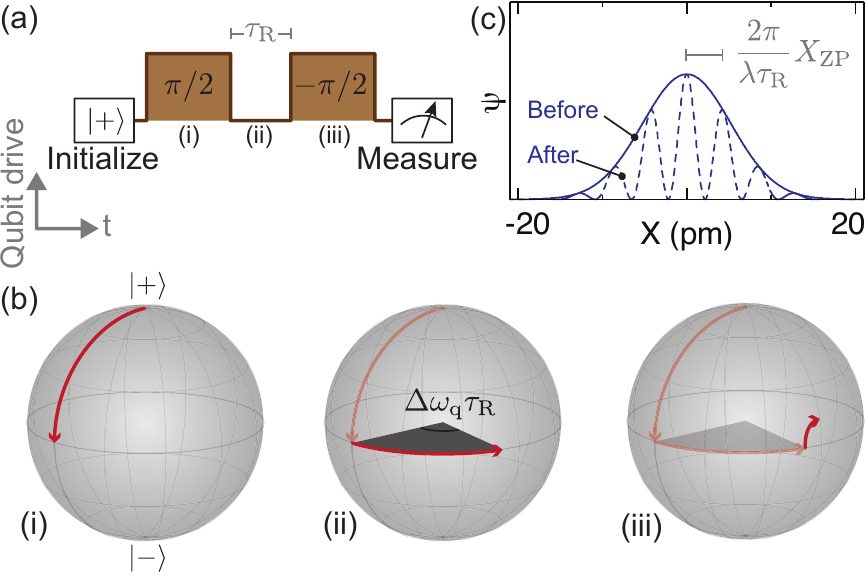}
	\caption{\label{Fig3} Using pulsed qubit operations to construct a diffraction grating. (a) Pulse sequence. (b) Evolution of the qubit on the Bloch sphere, during (i) preparation, (ii) precession at a rate set by the displacement-dependent qubit frequency shift $\Delta \omegaq$, and (iii) projection. (c) Effect of this pulse sequence on the ground-state mechanical wavefunction. Conditioning on the qubit measurement outcome $|+\rangle$, the \resonator wavefunction is multiplied by a periodic filter~(Eq.~\eqref{eq:gratingfilter}), analogous to a grating. Here we have taken $|\alpha|=1.9$.}
\end{figure}

The effective diffraction grating for the \resonator (Fig.~\ref{Fig3}) is implemented using Ramsey interferometry to generate a periodic spatial filter~\cite{Bennett2012, Asadian2014}. To understand how the grating arises, consider the time evolution operator $U(t)$ generated from Eq.~\eqref{eq:Hamiltonian}. As shown in Ref.~\cite{Asadian2014}, this time-ordered unitary is 
\begin{equation}
U(t) = \mathcal{R}(\omegam t)e^{-\frac{i\omegaq t}{2}\sigmaz}  \left( \mathcal{D}(\alpha)|-\rangle \langle -| +\mathcal{D}^\dagger(\alpha) |+\rangle \langle+| \right) ,
\label{eq:time_depend}
\end{equation}
where $\mathcal{D}(\alpha) = e^{\alpha\ad - \alpha^*a}$ and $\mathcal{R}(\theta) = e^{-i\theta \ad a}$ are \resonator displacement and rotation operators respectively, and
\begin{equation}
	\alpha = \frac{i}{2}\int_{-\infty}^t  e^{i\omegam t'}\lambda(t') dt'
	\label{eq:definealpha}
\end{equation}
is the amplitude of the coherent displacement. The superposition of displacement operators in Eq.~\eqref{eq:time_depend} applies equal and opposite momentum kicks (assuming $\mathrm{Re}(\alpha) = 0$ from here on~\footnote{If $\alpha$ is real, Eq.~(\ref{eq:time_depend}) leads to a position basis superposition, which is a $\pi/2$ rotation of the momentum superposition. In this way, any complex $\alpha$ can be understood in terms of a purely imaginary $\alpha$, followed by a rotation. Without loss of generality, and to keep the analogy to optical standing wave gratings, we therefore consider $\mathrm{Re}(\alpha) = 0$.})
to the \resonator depending on the state of the qubit. 
This is analogous to standing wave gratings in atom interferometry, which when decomposed into left and right propagating beams can be understood to impart superposed positive and negative impulses to atoms.

We now describe the protocol that realises this superposition of momentum kicks~(Fig.~\ref{Fig3}(a-b)). 
Following initialisation of the qubit in the excited state $|+\rangle$, a microwave burst applied to the qubit generates a $\pi/2$ rotation, preparing a superposition $(\ket{+} + \ket{-})/\sqrt{2}$. The mechanical interaction then causes the qubit state to precess at the displacement-dependent rate $\Delta\omegaq(t)=\frac{X(t)}{\XZP}\lambda(t)$. After an interval $\tau_\mathrm{R}$, a second $\pi/2$ burst is applied, followed by a $\sigmaz$ measurement. Conditioning on the qubit outcome gives a measurement operator that acts on the mechanical system, 
\begin{eqnarray}
 \Upsilon_\pm(\phi) &\equiv& \bra{\pm} \Pi^\dagger_\phi U(t)  \Pi_0\ket{+}\nonumber \\
 &=& \frac{ \mathcal{R}(\omegam t)}{2}[\mathcal{D}^\dagger(\alpha) \pm e^{i\phi +i\omegaq t}\mathcal{D}(\alpha) ].
 \label{eq:Upsilon1}
\end{eqnarray} 
Here $\Pi_\phi$ denotes a $\pi/2$ qubit rotation with phase $\phi$, and $\pm$ is the result of the $\sigmaz$ measurement. The (un-normalized) state of the \resonator after the interaction is,
\begin{equation}
\ket{\psi}_\mathrm{M}   \rightarrow \Upsilon_\pm \ket{\psi}_\mathrm{M} =  \mathcal{R}(\omegam t)\begin{array}{l}
\cos\\
\sin
\end{array}
\left(|\alpha|\frac{ X}{\XZP} + \frac{\phi}{2}\right)\ket{\psi}_\mathrm{M} ,
\label{eq:gratingfilter}
\end{equation}
where the $\cos(\cdot)$ or $\sin(\cdot)$ correspond to finding the qubit in the excited or ground state respectively~\footnote{Eq.~\eqref{eq:gratingfilter} can be inventively understood from Eq.~\eqref{eq:Upsilon1} by noting that for $\mathrm{Re(\alpha = 0)}$, $\mathcal{D}(\alpha) = e^{i|\alpha|(a + \ad)}$ and decomposing the exponentials in Eq.~\eqref{eq:Upsilon1} into trigonometric functions.}. The \resonator wavefunction is thus projected onto an effective diffraction grating with pitch $\pi \XZP /|\alpha|$ (Fig.~\ref{Fig3}(c)). Since the only difference between conditioning on the $\ket{\pm}$ outcomes is a relative change in phase of the effective grating, either measurement outcome may be used to define the grating. We refer to the Ramsey sequence followed by conditioning on the qubit measurement outcome as a grating operation. Its effect is to split the \resonator wavefunction into a superposition of left-moving and right-moving branches.

\label{sec:toggling}
A well-separated superposition, with both branches displaced by more than the zero-point amplitude, requires $|\alpha| \gtrsim 1$. To achieve this with our parameters, we require that $\lambda(t)$ is modulated at the mechanical frequency:
\begin{equation}
\lambda(t)=\lambda_0 g(t) \cos \Omega t ,
\label{eq:toggling}
\end{equation}
where $g(t)$ is a Gaussian envelope function with a maximum of unity and a full width at half maximum of $\tau_\lambda \gg 1/\Omega$.
Equation~\eqref{eq:definealpha} then gives
\begin{equation}
\alpha = i\frac{\sqrt{\pi}}{8\sqrt{\ln 2}} \lambda_0 \tau_\lambda.
\end{equation}
In the following, we take $\alpha \approx 1.9i$, thus achieving the desired momentum separation. A price to pay for this modulation is that the qubit and \resonator are susceptible to decoherence over the full duration of the envelope. With our parameters, we require $\tau_\lambda \approx 130$~ns, corresponding to $N\approx 17$ mechanical periods. This interaction time is short enough that the evolution of the \resonator-qubit system is well approximated as unitary. (See Appendix~\ref{app.1} for modelling of qubit dephasing and mechanical decoherence).

\subsection{Nanomechanical interferometry}

\begin{figure}
\includegraphics[width=86mm]{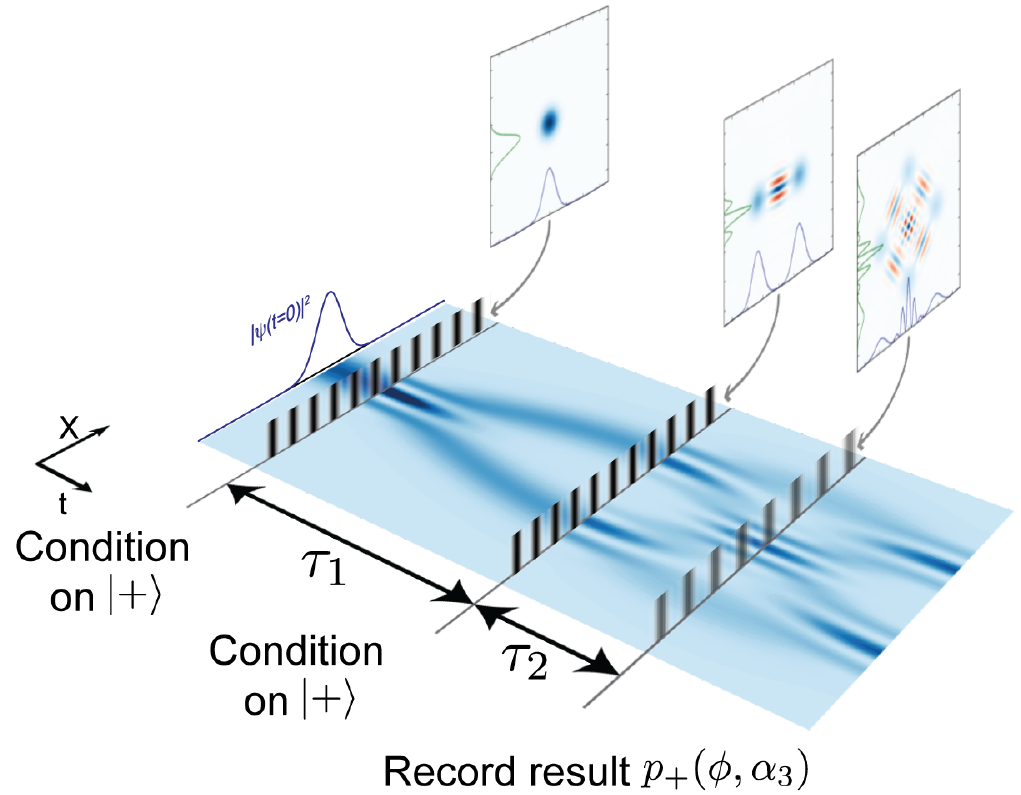}
\caption{\label{Fig4} Protocol for detecting nanomechanical interference. Following cooling (not shown), the first grating operation, with amplitude $\alpha_1$, diffracts the \resonator wavefunction into a superposition of left-moving and right-moving components. After an interval $\tau_1$ of free evolution, a second grating operation, with amplitude $\alpha_2$,
leads to recombination of the two components a quarter mechanical period later. To measure the resulting interference, a third (unconditioned) Ramsey sequence is applied after time $\tau_2$. The resulting qubit return probability $\prob(\alpha_3, \phi)$ probes the mechanical interference fringes. The main panel shows a simulated \resonator spatial density $|\psi(X)|^2$, beginning from the ground state, plotted as a function of displacement and time, with the grating operations and the final Ramsey measurement operation indicated schematically as filters. The \resonator Wigner distributions and marginals (see Fig.~\ref{Fig5}) are shown as insets just before each filter. To illustrate the continuing periodic evolution of the \resonator wavefunction, the spatial density beyond the final Ramsey measurement is plotted as it would be probed by applying the measurement instead at a later time.}
\end{figure}

We now show how a sequence of grating operations can be combined to create an interferometer~(Fig.~\ref{Fig4}).
The effect of a single grating operation (Eq.~\ref{eq:gratingfilter}) with phase $\phi = 0$ and amplitude $\alpha=\alpha_1$ is to divide the resonator's wavefunction into two components with added momentum $\pm |\alpha_1|\hbar/\XZP$. A second grating operation with the same amplitude and phase, applied after a duration $\tau_1=\pi/2\Omega$ corresponding to a quarter period of free evolution,
further splits the branches of the superposition, allowing quantum interference between recombined branches to be observed.
After a second evolution time $\tau_2$, the interference can be detected using a third Ramsey sequence.
In this step, there is no conditioning; the probability $\prob(\phi,\alpha_3)$ for the qubit to return to state $\ket{+}$ is measured as a function of the phase $\phi$ and amplitude $\alpha_3$ of the Ramsey sequence.
This probability is
\begin{eqnarray}
\prob(\phi,\alpha_3) &=& \mathrm{Tr}[\Upsilon^\dagger_+\Upsilon_+\rhoM] \nonumber \\
&=& \int dX' \cos^2\left( |\alpha_3|\frac{X'}{\XZP} + \frac{\phi}{2}\right) P(X'),
\label{eq:Pphi}
\end{eqnarray}
where $\rhoM$ is the density matrix describing the state of the \resonator immediately before the third grating, with position probability distribution $P(X)$.

Scanning the phase of the third Ramsey sequence is analogous to scanning the position of the third grating in a molecular interferometer~\cite{Eibenberger2013}, and the signature of interference is a sinusoidal dependence on $\phi$.
In fact, Eq.~(\ref{eq:Pphi}) can be understood as a Fourier decomposition in which each choice of $|\alpha_3|$ probes the component of  $P(X)$ with wave number $2|\alpha_3|$. From here on we will use $x \equiv X/\XZP = (a + \ad)$ as a dimensionless position operator.

\begin{figure}
	\includegraphics[width=86mm]{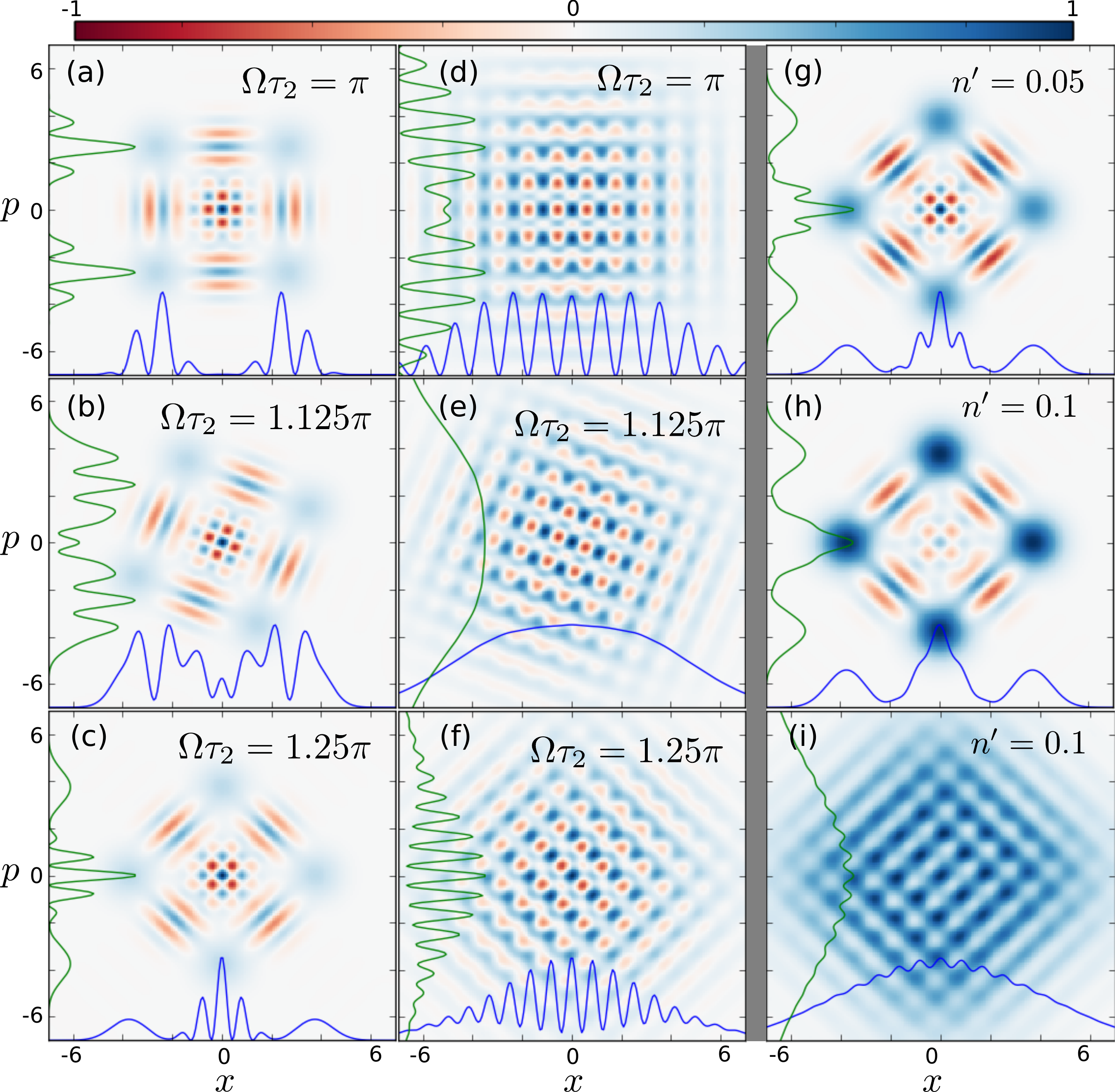}
	\caption{\label{Fig5} Wigner distributions of the resonator's state at different values of $\tau_2$ showing: (a)-(c) maximum interference from an initial pure state with unitary evolution, (d)-(f) loss of interference due to initial thermal phonon occupation $\bar{n} = 5$ (with unitary evolution), and (g)-(i) loss of interference with non-unitary evolution equivalent to adding 0.05 (g) or 0.1 (h)-(i) thermal phonons. The initial state for (g) and (h) is the vacuum state, and the initial state of (i) is a thermal state with $\bar{n} = 5$ phonons. Blue and green traces show the position and momentum marginals respectively, normalized to the same maximum. Each plot uses $|\alpha| \approx 1.9$ and the color scale has been normalised to a maximum of unity.}
\end{figure}

Our goal now is to use $\prob$ to distinguish \emph{quantum} interference from classical fringes that might appear in the resonator's probability distribution $P(x)$.
Classical fringes might arise, for example, from
the shadow of the diffraction gratings, or from Moir\'{e} patterns.
To recognise the quantum interference, we plot the \resonator Wigner distribution at different times $\tau_2$ after the second grating operation, choosing $|\alpha_1| = |\alpha_2| \approx 1.9$ (Fig.~\ref{Fig5}). The effect of applying the first grating operation is to vertically slice the distribution with $\cos^2(|\alpha|x)$, and to prepare a superposition of two momentum states (see second inset in Fig.~\ref{Fig4} plotting the state after the first grating, rotated by one quarter period). Since the second grating operation is applied a quarter period after the first, it slices along an orthogonal axis, leading to the ``compass-like'' distribution of Fig.~\ref{Fig5}(a)~\cite{Zurek2001}. 

The compass distribution is intuitively understood: the quarter-period rotation after the first grating turns the momentum superposition state into a position superposition. Each branch of the superposition then passes the grating, generating its own momentum superpositions and resulting in the four-lobe compass state. The compass state is clearly visible if the \resonator is initially prepared in the ground state, $\bar{n} = 0$ (Fig.~\ref{Fig5}(a-c)), but is washed out if the \resonator is initially in a thermal state, leaving only the orthogonally sliced pattern visible (Fig.~\ref{Fig5}(d-f)). During the evolution time $\tau_2$, the Wigner distribution rotates (Fig.~\ref{Fig5}(b-c) and (e-f)), so that the interference fringes oscillate between the position and momentum marginals (plotted in blue and green traces respectively). Interference patterns arise when two lobes of a coherent quantum superposition overlap in position space, for example the north-east and south-east lobes in Fig.~\ref{Fig5}(a) interfering around $x\approx 3$, or the north and south lobes in Fig.~\ref{Fig5}(c) interfering around $x\approx 0$. The wave numbers present in the position marginal (which is measured by the third grating) are proportional to the momentum separation of lobes in phase space, geometrically illustrating the $\sqrt{2}$ ratio between wave number components present in Fig.~\ref{Fig5}(a) and Fig.~\ref{Fig5}(c).

The interference fringes arising when the \resonator is initially prepared in its ground state can be compared with those arising from an initial thermal state ($\bar{n}=5$).  
If the width of the initial thermal state (as in Fig.~\ref{Fig5}(d)-(f)) is larger than the superposition size ($\sqrt{\bar{n}} > |\alpha|$), then the vertical slicing of the grating is no longer accompanied by a distinct momentum superposition, but rather an increase in the momentum variance (as seen by the broader than Gaussian position and momentum distributions in Figure.~\ref{Fig5}(e)), and results in a checker-board pattern.

We can now see the distinction between quantum and classical fringes appearing in the marginal distributions. The shadow of the gratings is dominated by components close to the wave number $2|\alpha_1|$ (Fig.~\ref{Fig5}(a),(d)). The Moir\'{e} patterns arising from the checkerboard have components close to at most two wave numbers, $2|\alpha_1|$ and $2\sqrt{2}|\alpha_1|$ (see Appendix~\ref{app.2}). By contrast, the quantum interference pattern (Fig.~\ref{Fig5}(a-c)) has multiple wave numbers components at each $\omegam \tau_2$, as seen in the position marginals. Furthermore, quantum interference appears for all evolution times $\tau_2$, whereas classical fringes are washed out (Fig.~\ref{Fig5}(e)) except at particular fractions of the mechanical period. Hence for this protocol, a marginal $P(x)$ with multiple wave number components, observed at all rotation angles $\omegam \tau_2$, indicates quantum interference.

We now show that this interferometer is a sensitive probe for quantum decoherence, which damps the interference fringes in $P(x)$, and therefore destroys the signature of quantum coherence in $\prob$. To model decoherence following the second grating operation, we consider weak thermalisation of the state, resulting in a decohered state (superscript d)
\begin{eqnarray}
\rhoM^{\mathrm{(d)}} = \int d^2\beta \frac{e^{-|\beta|^2/n'}}{\pi n'}\mathcal{D}(\beta)\rhoM \mathcal{D}^\dagger(\beta)
\label{eq:rho_loss}
\end{eqnarray}
where $n'$ is the number of thermal phonons effectively added to the \resonator, causing loss of quantum coherence. The loss of coherence is equivalent to convolving $P(x)$ with a Gaussian of width $\sqrt{n'}$, thereby exponentially damping oscillations of wave number $|\alpha|>\sqrt{n'}$~(Appendix~\ref{app.1}). Figure~\ref{Fig5} (g)-(i) plots the effect of loss of coherence between the second and third gratings, assuming an initial ground state (g) and (h), and an initial ($\bar{n} = 5$) thermal state (i). The decoherence is modelled only after the second grating, so that without thermalisation plots \ref{Fig5}(g)-(h) would coincide with \ref{Fig5}(c), and \ref{Fig5}(i) would coincide with \ref{Fig5}(f).  The plots show that even the addition of a fraction of a phonon drastically suppresses the interference pattern in $P(x)$ and the corresponding signature in $\prob$.

\begin{figure}
	\includegraphics[width=86mm]{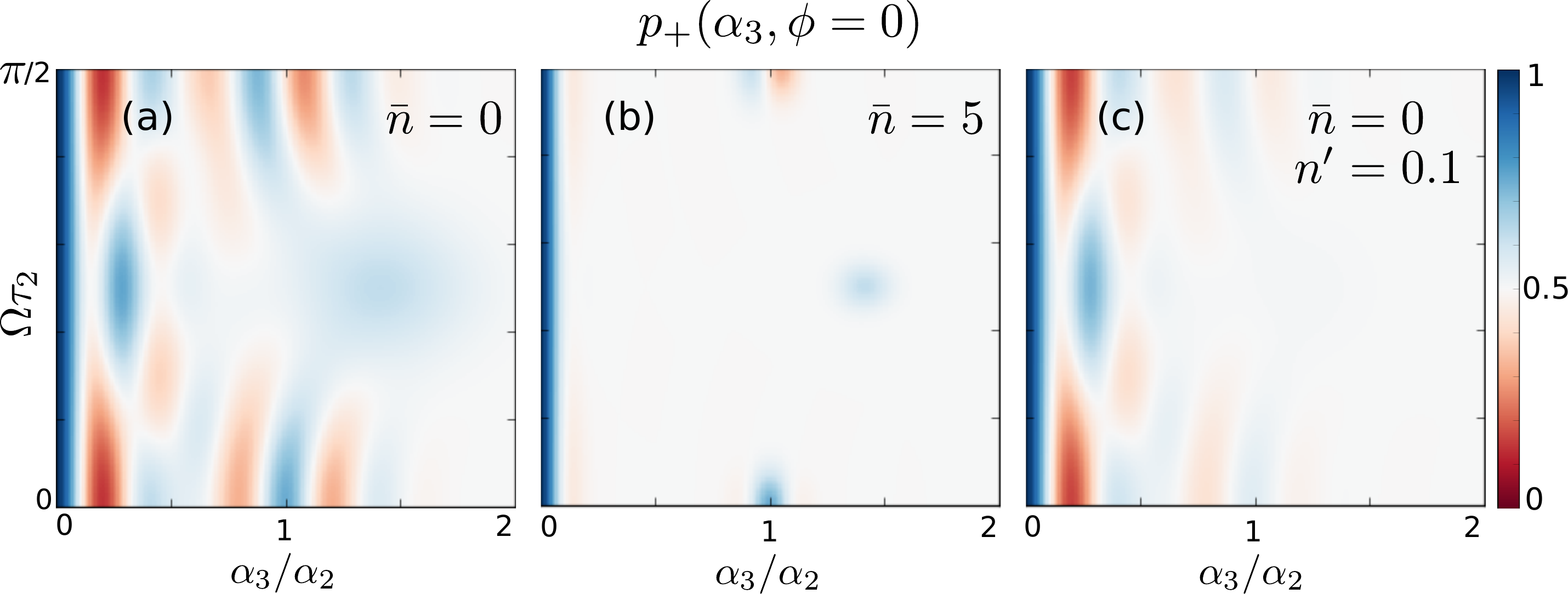}
	\caption{\label{Fig6} The probability to find the qubit in its excited state, $\prob (\alpha_3)$, can be used to probe different wave number components in the position probability distribution $P(x)$. As the \resonator evolves (increasing $\omegam \tau_2$), different wave numbers are present in $P(x)$ (see Fig.~\ref{Fig5}). The probability $\prob(\alpha_3)$ is plotted for the \resonator initially in (a) a pure state, (b) a thermal state of $\bar{n} = 5$ phonons, and (c) a pure state, but with decoherence (represented by $n'$ added phonons) after the second grating.}
\end{figure}

Finally in Fig.~\ref{Fig6} we show explicitly how these two effects - 
 thermal occupation before the inteferometry, and decoherence during the interferometry - 
degrade the quantum signatures  in $\prob$. 
In the ideal case (Fig.~\ref{Fig6}(a)), there are several values of $\alpha_3$ at each time $\tau_2$ that give non-trivial probabilities of $\prob$. In contrast, beginning in a thermal state with $\bar{n} = 5$ phonons (Fig.~\ref{Fig6}(b)), all fringes are washed out, with the exception of the shadow of the grating (at $\alpha_3/\alpha_2 = 1$ and $\omegam \tau_2 = 0, \pi/2$) and the Moir\'{e} pattern (at $\alpha_3/\alpha_2 = \sqrt{2}$ and $\omegam\tau_2 = \pi/4$). 
Loss of coherence during the interferometry (Fig.~\ref{Fig6}(c)) leads to a qualitatively different behavior, namely damping of all features in $\prob$, including the classical fringes, for $|\alpha_3|>\sqrt{n'}$. This mechanical interferometery scheme could thus be used as a specific probe of quantum decoherence during the mechanical evolution.

\section{Conclusion}
We have introduced the displacemon electromechanical system that provides sufficiently strong coupling to generate and detect quantum interference of a massive object containing a quarter of a million atoms. By considering device parameters based on current technology, our qubit-\resonator displacemon can achieve effective ultrastrong coupling using a modulated coupling scheme. A similar scheme could also be applied to other kinds of solid-state qubits coupled to high-quality resonators, such as spin qubits coupled to nanotubes~\cite{Ohm2012,Palyi2012}, diamond defects coupled to cantilevers~\cite{Kolkowitz2012}, or piezoelectric \resonators coupled to superconducting qubits~\cite{Manenti2017,Chu2017}. However, the parameters estimated for the
proposed device of Fig.~\ref{Fig1} may be particularly experimentally favourable for this implementation, because they imply that the coupling exceeds both the thermal decay rate of the resonator and the typical dephasing rate of a qubit~(see Appendix~\ref{app.3}). Importantly, our scheme does not rely on degeneracy between the qubit and the \resonator~\cite{OConnell2010}, nor on qubit coherence over the lifetime of the mechanical superposition~\cite{Wan2016}.

Using the effective ultrastrong coupling we have shown that it is possible to extend matter wave interferometry to nano-mechanical \resonators, opening up a range of new devices that can be used to study quantum physics at the meso-scale. Furthermore, interferometry  performed on the wavefunction of a mechanically bound \resonator is qualitatively distinct from existing free-particle interferometry techniques. An important advantage of nanomechanical \resonators for quantum tests is that they can readily be extended to probe much larger masses than can be accessed in molecular vapours or even levitated nanoparticles. Although the nanotube \resonator considered here does not have enough mass to seriously challenge the interesting parameter regime of extrinsic collapse theories, a similar protocol could be extended to more massive objects still well within the range of nanomechanics. This research direction could allow for testing specific theories of quantum collapse~\cite{Bassi2013}, as an alternative to proposals based on single-photon optomechanics~\cite{Marshall2003}, or levitated nanoparticles~\cite{Bateman2013,Wan2016}. Finally, multiple \resonators coupled to the same qubit (such as the pair of nanotube junctions in Fig.~\ref{Fig1}) could allow for implementation of entanglement generation between massive objects, leading to Bell tests of mechanical \resonators~\cite{Hofer2016}.

\section{Acknowledgements}
We thank P. J.~Leek and E. M.~Gauger for discussions. We acknowledge FQXi, EPSRC (EP/N014995/1), and the Royal Academy of Engineering. KK would like to thank the Department of Materials, University of Oxford for their hospitality during the initial stages of this work.

\appendix
\section{Device parameters} \label{app:parameters}

We assume device parameters based on a mixture of experiment and simulation as follows.
We take the parameters of the nanotube and the junctions from the nanotube SQUID device of Ref.~\cite{Schneider2012}. For a \resonator with length $l=800$~nm, the frequency was measured as $\omegam/2\pi = 125$~MHz, which with estimated diameter $D = 2.5$~nm and mass $m=5 \times 10^{-21}~\mathrm{kg} = 3 \times 10^{6}$ amu leads to $\XZP=3.7$~pm, typical of nanotube \resonators. In each SQUID junction, a critical current $\Ic \approx 12$~nA was achieved, implying $\EJ^0/h = 12$~GHz. 

For the qubit, the charging energy is set by the electrode geometry, which is a design choice. Finite-element capacitance simulation for the device of Fig.~\ref{Fig1}(b), with qubit diameter $340~\mu$m, gives $\EC/h = 0.2$~GHz, typical of qubit devices and well into the transmon regime $\EJ^0 \gg \EC$~\cite{Schreier2008}. The maximal qubit frequency is then $\omegaq^0/2\pi = \sqrt{8\EJ\EC} / h = 4.4$~GHz, and the calculated qubit energy levels are shown in Fig.~\ref{Fig1}(c).

For the in-plane magnetic field we assume $B_{||} = 0.5$~T, which nanotube SQUIDs can withstand~\cite{Schneider2012}. The operating flux point should then be chosen to maximise $\lambda$, while still maintaining a qubit frequency compatible with microwave \resonators. We assume flux bias $\Delta\Phi/\Phi_0 = -0.84$, leading to a qubit frequency $\omegaq = 2\pi \times 2.19$~GHz~$=\omegaq^0/2$ (dashed vertical line in Fig.~\ref{Fig1}(c-d)). With symmetric junctions, and assuming that the restoring force on the nanotube is dominated by tension, the coupling is then $\lambda/2\pi=8.5$~MHz. Since the coupling can be reduced by tuning $\Delta \Phi$ towards zero, we take this as the maximum coupling strength $\lambda_0$.

In a realistic device, we must take account of asymmetry between the junctions. Denoting the two critical currents by $I_{c1}$ and $I_{c2}$, with $\delta \equiv 2(I_{c1}-I_{c2})/(I_{c1}+I_{c2})$ being the asymmetry parameter, we have $\EJ=\EJ^0 \sqrt{\cos^2 (\pi \Delta\Phi/2\Phi_0)+\frac{\delta^2}{4}\sin^2 (\pi \Delta\Phi/2\Phi_0)}$ with a corresponding modification to Eq.~\eqref{eq:domegadX}~\cite{Koch2007,Shevchuk2017}. This asymmetry leads to a small reduction in $\lambda_0$ (Fig.~\ref{Fig1}(c)-(d)).

For these parameters, the device would be in the strong coupling regime ($\lambda_0 > \kB T/\hbar\Qm, 1/T_2$) for comparatively modest \resonator quality factor $\Qm \gtrsim 15$ and $T_2 \gtrsim 120$~ns. To access the ultrastrong coupling regime ($\lambda_0>\omegam$) is more challenging, but may be possible~\cite{Shevchuk2017}: if the suspended length could be increased to  $l\approx1.6~\mu$m and the tension reduced to zero while keeping other parameters unchanged, the coupling would be $\lambda_0/2\pi \approx \omegam/2\pi \approx 25$~MHz.
However, in the simulations we do not make this assumption, but instead assume that effective ultrastrong coupling is engineered by toggling $\lambda(t)$ as in Eq.~\eqref{eq:toggling}.

\section{Decoherence}\label{app.1}

Here we model the effect of decoherence on the interference. For our chosen parameters, the effect is estimated to be weak, because the interaction time $\tau_\lambda \approx 130$~ns is short compared with other timescales. For a superconducting qubit, the decoherence time is typically $T_2 > 1~\mu\mathrm{s} \gg \tau_\lambda$, so we may neglect qubit dephasing. For the \resonator, the high quality factor $\Qm$ suppresses thermal decoherence; assuming $\Qm = 10^{5}$, there are $\bar{n}/\Qm \approx 5\times 10^{-5}$ phonons exchanged with the thermal environment every \resonator period, or 1 phonon exchanged every $\sim 10^3$ realisations of the interaction. Below, we model the effect of decoherence in detail.

\subsection{Qubit dephasing}

We model dephasing by adding a stochastic frequency shift to the qubit, changing the Hamiltonian (Eq.~\eqref{eq:Hamiltonian}) to 
\begin{equation}
H =\hbar \omegam \ad a + \frac12\hbar\omegaq \sigmaz + \frac12\hbar\lambda(t) (a + \ad) \sigmaz + \hbar\sqrt{\gamma/2}\;\xi(t)\sigmaz ,
\label{eq:H_dec}
\end{equation}
where $\gamma$ is the qubit dephasing rate and $\xi(t)$ is a delta-correlated white noise term satisfying $\mathcal{E}(\xi(t)) = 0$ and $\mathcal{E}(\xi(t)\xi(t')) = \delta(t-t')$. Here $\mathcal{E}(\cdot)$ denotes an average over realisations of this stochastic process. Moving into the interaction picture, the unitary generated by this Hamiltonian is 
\begin{equation}
 U(t) = e^{-iW(t)\sqrt{\gamma/2}\sigmaz}  \left( \mathcal{D}(\alpha)|-\rangle \langle -| +\mathcal{D}^\dagger(\alpha) |+\rangle \langle+| \right) ,
 \label{eq:u2}
\end{equation}
where $W(t) = \int_{t_0}^t \xi(t') dt'$ is a stochastic variable with a mean of zero and a variance of $t$. Since there is classical uncertainty in the realisation of $W(t)$, the joint state of the \resonator-qubit system will be mixed. Due to this classical uncertainty, the measurement operator cannot be understood as mapping pure states to pure states  as assumed in Eq.~\eqref{eq:Upsilon1}.

We must therefore consider the measurement procedure (used to impose the grating) in the density matrix description. Before switching on the interaction, i.e while $\lambda(t) = 0$, the $\pi/2$ pulse changes the $\ket{+}\bra{+}$ state of the qubit to $\rho_q =\frac14( \ket{+} +\ket{-})(\bra{+}+\bra{-})$. The state of the mechanical \resonator is left unchanged in an arbitrary state $\rhoM$. This joint state must be separable, because the initialization of the qubit state at the beginning of the grating operation has the effect of destroying any qubit-\resonator entanglement.

As the interaction is switched on, the joint state of the system evolves according to
\begin{eqnarray}
\rho(t, W(t)) &=& U(t)\rho_q\otimes \rhoM U^\dagger(t) \nonumber \\	
 &=& \frac{1}{2}e^{2i\sqrt{\gamma/2}W(t)}\ket{-}\bra{+}\otimes \mathcal{D}(\alpha)\rhoM \mathcal{D}(\alpha) \nonumber \\
 & & + \frac{1}{2}e^{-2i\sqrt{\gamma/2}W(t)}\ket{+}\bra{-}\otimes \mathcal{D}^\dagger(\alpha)\rhoM \mathcal{D}^\dagger(\alpha) \nonumber \\
 & & + \frac{1}{2}\ket{+}\bra{+}\otimes \mathcal{D}^\dagger(\alpha)\rhoM \mathcal{D}(\alpha) \nonumber \\
 & & + \frac{1}{2}\ket{-}\bra{-}\otimes \mathcal{D}(\alpha)\rhoM \mathcal{D}^\dagger(\alpha),
\end{eqnarray}
\\
where $U(t)$ is the unitary operator in Eq.~(\ref{eq:u2}). Since $W(t)$ is unknown, the resulting quantum state at time $t$ must be weighted by the probability of obtaining a particular realisation of $W(t)$, where $P(W(t)) = \exp[-W^2(t)/2t]/\sqrt{2\pi t}$, 
\begin{eqnarray}
\rho(t) &=& \int_{-\infty}^{\infty} \rho(t, W(t)) P(W(t)) dW(t). 
\end{eqnarray}
Projecting the qubit onto the state $(\ket{+} + e^{-i\phi}\ket{-})/\sqrt{2}$ gives the unnormalised conditional state of the mechanical \resonator
\begin{eqnarray}
\rho_{\mathrm{m},\pm} &\sim& \frac12(\bra{+} \pm e^{i\phi}\bra{-})\rho(t)(\ket{+} \pm e^{-i\phi}\ket{-})\\
&\sim& 
\frac14 \left[\mathcal{D}^\dagger(\alpha)\rhoM \mathcal{D}(\alpha) + \mathcal{D}(\alpha)\rhoM\mathcal{D}^\dagger(\alpha)\right. \nonumber \\
& & \left.\pm e^{-\gamma t}\left(e^{-i\phi} \mathcal{D}^\dagger(\alpha)\rhoM \mathcal{D}^\dagger(\alpha) + e^{i\phi}\mathcal{D}(\alpha)\rhoM \mathcal{D}(\alpha)\right)\right] \nonumber \\
\label{eq:rhoc}
\end{eqnarray}
where we have used $\sim$ because the right hand side is unnormalised. Separating this into coherent and incoherent terms, we find
\begin{equation}
\begin{split}
\rho_{\mathrm{m},\pm} \sim &\frac{ e^{-\gamma t}}{4}[\mathcal{D}^\dagger(\alpha) \pm e^{i\phi}\mathcal{D}(\alpha) ] \rhoM [\mathcal{D}(\alpha) \pm e^{-i\phi}\mathcal{D}^\dagger(\alpha) ] \\
&+ \frac{1 - e^{-\gamma t}}{4}\left[\mathcal{D}^\dagger(\alpha)\rhoM \mathcal{D}(\alpha) + \mathcal{D}(\alpha)\rhoM\mathcal{D}^\dagger(\alpha)\right].
\end{split}
\label{eq:deph}
\end{equation}
We notice that the first term is proportional to $\Upsilon_\pm \rhoM\Upsilon^\dagger_\pm$ where $\Upsilon_\pm$ is given in Eq.~\eqref{eq:Upsilon1} (with $\omegaq  = 0$). The first term in Eq.~\eqref{eq:deph} is exactly the state that one would expect if the grating protocol worked perfectly, while the second term is an incoherent mixture of displacements. We can therefore understand qubit dephasing as some classical probability that the \resonator coherently passed the grating, and some probability that we ended up with an incoherent mixture. Since the normalisation is state dependent, we cannot simply relate the coefficients in Eq.~\eqref{eq:deph} with direct probabilities. However we can say the relative probability of introducing an incoherent mixture is $(1-e^{-\gamma t})/e^{-\gamma t} = e^{\gamma t}-1 \approx \gamma t$ for short times, or low dephasing rate. The trace of $\rho_{\mathrm{m},\pm}$ is the probability of finding the qubit in the $\ket{\pm}$ state and using Eq.~\eqref{eq:rhoc} we may read off,
\begin{equation}
p_\pm(\alpha,\phi)   = \frac{1}{2} \pm \frac{e^{-\gamma t}}{4} [e^{i\phi}\chi(-2\alpha) + e^{-i\phi}\chi(2\alpha) ],
\label{eq:P_k}
\end{equation}
where $\chi(\cdot) = \mathrm{Tr}[\mathcal{D}(\cdot)\rho]$ is the characteristic function of the mechanical state (using $\mathrm{Tr}[\mathcal{D}(\alpha)\rho\mathcal{D}(\alpha)] = \mathrm{Tr}[\mathcal{D}(\alpha)\mathcal{D}(\alpha)\rho] = \mathrm{Tr}[\mathcal{D}(2\alpha)\rho]$, etc.). The characteristic function is related to the Wigner distribution via a symplectic Fourier transform, 
\begin{equation}
\chi(\alpha) = \int dx dp W(x,p) e^{ix\alpha_i - ip\alpha_r}  
\end{equation}
where $\alpha_{r(i)}$ is the real (imaginary) part of $\alpha$. If Re$(\alpha) = 0$, then  
\begin{eqnarray}
\chi(2\alpha_i) &=& \int dx dp W(x,p) e^{2ix\alpha_i} \nonumber \\
&=& \int dx P(x) [\cos(2\alpha_i x) + i \sin(2\alpha_ix)],
\end{eqnarray}
where the complex part must vanish as $\chi(\cdot)$ is a real function (for states with $\pi$ rotational symmetry). This is simply the overlap integral between the position probability distribution $P(x)$ and a diffraction grating with a pitch $|\alpha|$. Therefore
\begin{equation}
p_\pm(\alpha, \phi) = \frac12 \pm \frac{e^{-\gamma t}\int dx P(x) \cos(2 x|\alpha|)}{2}\cos(\phi), 
\end{equation}
which is exactly probing the $2|\alpha|$ wave number components in $P(x)$, with a reduced amplitude from the qubit dephasing. 

Thus we have seen the effect of qubit dephasing is to introduce some probability of having an incoherent mixture of different momentum kicks, thus suppressing any signatures of interference in the outcomes of qubit measurements. Since the duration of the protocol is on the order $\sim N$ mechanical oscillations, $t\approx 2\pi\times N/\omegam$, to neglect qubit dephasing requires $\gamma/\omegam\ll 1/N$. For the parameters discussed in the main text, this requires $\gamma/2\pi <$~1~MHz. 

\subsection{Loss of \resonator coherence}
To see that oscillations in $\prob(|\alpha|, \phi)$ are a quantum effect, we consider the effect of adding $n'$ thermal phonons to the state of the \resonator immediately before the third grating (Eq.~\eqref{eq:rho_loss}, restated here for convenience), 

\begin{equation}
\rho_{\mathrm{m}}^{\mathrm{(d)}} = \int d^2 \beta \frac{e^{-|\beta|^2/n'}}{\pi n'} \mathcal{D}[\beta]\rhoM \mathcal{D}^\dagger[\beta].
\end{equation}
In this case
\begin{eqnarray}
\prob(|\alpha|,\phi) &=& \Trm[\Upsilon^\dagger_+ \Upsilon_+ \rho_{\mathrm{m}}^{\mathrm{(d)}} ] \nonumber\\ 
&=& \int d^2\beta  \frac{e^{-|\beta|^2/n'}}{\pi n'} \Trm[\mathcal{D}^\dagger[\beta]\Upsilon^\dagger_+ \Upsilon_+ \mathcal{D}[\beta]\rhoM] \nonumber \\
&=& \int dx'd\beta_r \frac{e^{-\beta_r^2/n'}}{\sqrt{\pi n'}} P(x') \nonumber\\
& & ~~~~ \times \cos^2\left(|\alpha|(x' + 2\beta_r) +\frac{\phi}{2}\right) \nonumber \\
&=& \frac{1}{2} + \frac{e^{-4 n'|\alpha|^2}\int dx P(x) \cos(2 x|\alpha|)}{2}\cos(\phi), \nonumber \\
\end{eqnarray}
where $\Trm$ denotes a trace over the mechanical degrees of freedom. We therefore see that any loss of coherence between the second and third grating reduces the amplitude of the oscillations in $\prob$ by a factor $e^{-4n'|\alpha^2|}$.

\begin{figure}
	\includegraphics[width=86mm]{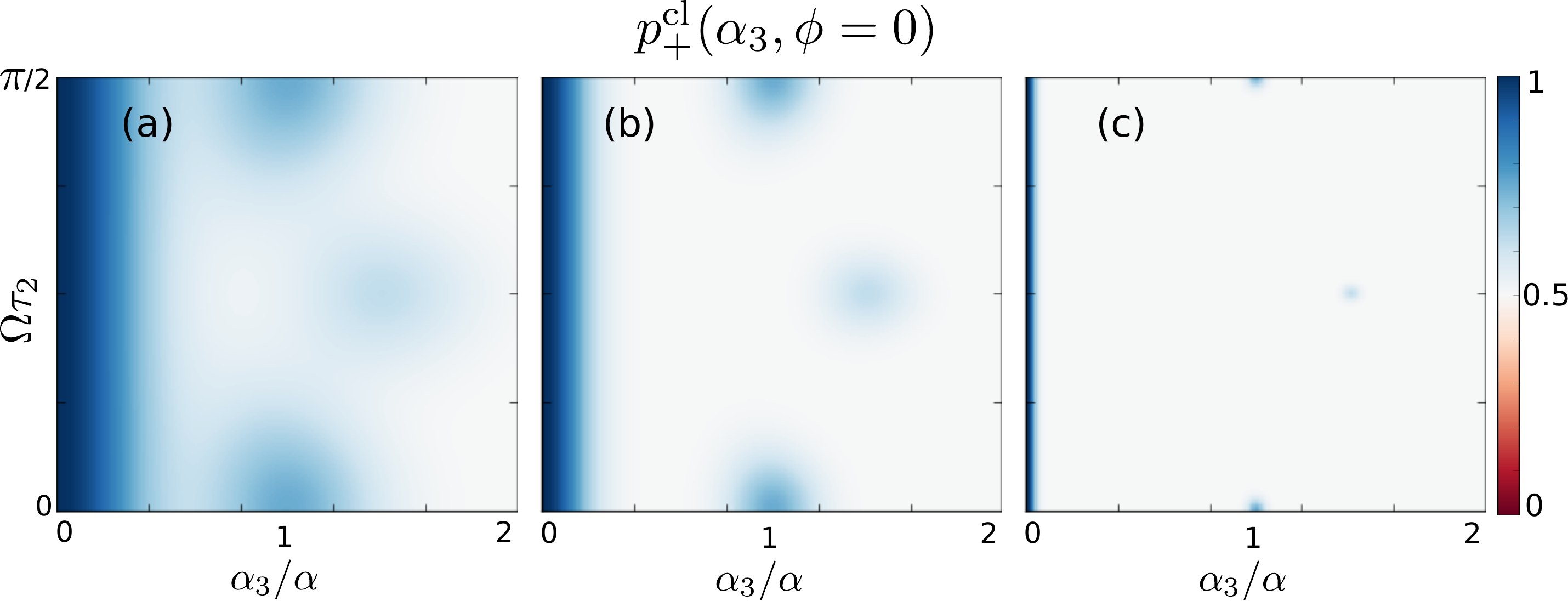}
	\caption{\label{Fig7} Probability $\prob$ of finding the qubit in the excited state if the \resonator were described by a classical probability distribution using width $\sigma = 0.5, 1, 5$ from (a) to (c).}
\end{figure}

\section{Classical interference patterns}\label{app.2}
To verify that oscillations are in fact due to quantum interference, we will consider what types of interference patterns can be understood using a classically description of the \resonator. Consider a classical (superscript cl) checkerboard phase space probability density of dimensionless variables $(x,p)$, 
\begin{equation}
P^{\mathrm{cl}}(x,p) = \frac{1}{N}\exp\left[-\frac{x^2 + p^2}{2\sigma^2}\right]\cos^2\left( \alpha x\right)\cos^2\left(\alpha p\right),
\end{equation}
where $N$ is a normalization factor and $\sigma$ and $\alpha$ are the width and wave number of the checkerboard pattern respectively. This is the conditional state after the \resonator has passed the first two gratings (of $\alpha = \alpha_1 = \alpha_2$) separated by a quarter period rotation. As the distribution rotates in phase space, we are interested in the wave number components present in the reduced probability of $P^{\mathrm{cl}}(x)$ where
\begin{equation}
P^{\mathrm{cl}}(x) = \int  P^{\mathrm{cl}}(x\cos\theta + p\sin\theta, p\cos\theta - x\sin\theta) dp. 
\label{eq:fourier}
\end{equation}
and $\theta$ is the phase space rotation angle. If this probability distribution is now probed by a third grating, then the probability of finding the qubit in the $\ket{+}$ state is (from Eq.~\eqref{eq:Pphi})
\begin{eqnarray}
\prob^{\mathrm{cl}}(\alpha_3, \phi) &=&  \int dx' \cos^2\left( |\alpha_3|x' + \frac{\phi}{2}\right) P^{\mathrm{cl}}(x')\nonumber \\
&\propto& e^{ -2 \sigma ^2 \left(|\alpha_3|^2+2 |\alpha_3| |\alpha |(\sin (\theta )+\cos (\theta ))+2 |\alpha| ^2\right)} \nonumber \\
& & \times \left[e^{8 |\alpha_3| |\alpha|  \sigma ^2 \cos (\theta )}+2 e^{2 |\alpha|  \sigma ^2 (2 |\alpha_3| \cos (\theta )+|\alpha| )}+1 \right] \nonumber \\
& & \times \left[e^{8  |\alpha_3|  |\alpha|  \sigma ^2 \sin (\theta )}+2 e^{2  |\alpha|  \sigma ^2 (2  |\alpha_3| \sin (\theta )+ |\alpha| )}+1 \right] \nonumber\\
& & \times \cos(\phi)
\end{eqnarray}
the amplitude of which peaks at $\{\omegam \tau_2, |\alpha_3/\alpha|\} = \{\frac12 n\pi, 1\}$ and $\{(\frac12 n +\frac14)\pi, \sqrt{2}\}$ for $\sigma \geq  |\alpha|$. A plot of this function (Fig.~\ref{Fig7}) looks qualitatively the same as Fig.~\ref{Fig6}(b), with the difference being attributable to the superposition of momentum kicks that accompanies the measurement when the \resonator is quantized. This confirms that the probability of finding the \resonator in the $\ket{+}$ state can therefore be used to distinguish quantum interference patterns (Fig.~\ref{Fig6}(a) and(c)) from classical Moir\'{e} patterns that arise from a classical probability distribution (Fig.~\ref{Fig7}).

\section{Other device implementations}\label{app.3}
\begin{table*}
	\begin{tabular}{lrrrrrrrr}
	\hspace{4cm}
		&	\hspace{1.55cm}	&	\hspace{1.55cm}	&	\hspace{1.55cm}	&	\hspace{1.55cm}	&	\hspace{1.55cm}	&	\hspace{1.55cm}	&	\hspace{1.55cm}	&	\hspace{1.5cm}	\\
	\hline
	\hline
	System 											
		&	\multicolumn{1}{c}1	&	\multicolumn{1}{c}2	&	\multicolumn{1}{c}3	&	\multicolumn{1}{c}4	&	\multicolumn{1}{c}5	&	\multicolumn{1}{c}6	&	\multicolumn{1}{c}7	&	\multicolumn{1}{c}8	\\
	Reference
	&	\multicolumn{1}{c}{This work}	&	\multicolumn{1}{c}{\cite{Scala2013}}	&	\multicolumn{1}{c}{\cite{Kolkowitz2012}}	&	\multicolumn{1}{c}{\cite{Kolkowitz2012}}	&	\multicolumn{1}{c}{\cite{Meesala2016}}	&	\multicolumn{1}{c}{\cite{Braumuller2016,Teufel2009}}	&	\multicolumn{1}{c}{\cite{Braumuller2016,Kolkowitz2012}}	&	\multicolumn{1}{c}{\cite{Braumuller2016,Yuan2015}}	\\
	\hline
\hline
Parameters \\
\hline
	Frequency $\omegam/2\pi$ (Hz)		
		&	$1.25 \times 10^8$	&	$10^5$	&	$8.0 \times 10^4$	&	$10^6$	&	$9.19 \times 10^5$	&	$1.0 \times 10^6$	&	$1.23 \times 10^5$	&	$10^6$	\\
	Mass $m$ (amu)									
		&	$3 \times 10^6$	&	$8 \times 10^9$	&	$7 \times 10^{15}$	&	$7 \times 10^{15}$	&	$3 \times 10^{10}$	&	$7 \times 10^{12}$	&	$1 \times 10^{17}$	&	$7 \times 10^{15}$	\\
	Quality factor $\Qm$					
		&	$10^{5}$	&	-	&	$2 \times 10^{2}$	&	$10^{6}$	&	$1 \times 10^{4}$	&	$6 \times 10^{5}$	&	$4 \times 10^{7}$	&	$10^{6}$	\\
	Temperature $T$ (K)								
		&	$3.3 \times 10^{-2}$	&	$1 \times 10^{-3}$	&	$3 \times 10^{2}$	&	$1 \times 10^{-1}$	&	$3 \times 10^{2}$	&	$1.3 \times 10^{-1}$	&	$1.8 \times 10^{-1}$	&	$1 \times 10^{-1}$	\\
	Qubit $T_2^*$									
		&	$2 \times 10^{-6}$	&	$10^{-4}$	&	$1.6 \times 10^{-4}$	&	$10^{-3}$	&	$2 \times 10^{-6}$	&	$2 \times 10^{-6}$	&	$2 \times 10^{-6}$	&	$2 \times 10^{-6}$	\\
	Coupling constant $\lambda/2\pi$ (Hz)			
		&	$8.5 \times 10^{6}$	&	$2 \times 10^{4}$	&	$1.6 \times 10^{1}$	&	$2 \times 10^{4}$	&	$1.8$	&	$7 \times 10^{4}$	&	$3$	&	$2$	\\
\hline
Dimensionless figures of merit \\
\hline
	Coupling parameter $\lambda/\omegam$
		&	$7 \times 10^{-2}$	&	$2 \times 10^{-1}$	&	$2 \times 10^{-4}$	&	$2 \times 10^{-2}$	&	$2 \times 10^{-6}$	&	$6 \times 10^{-2}$	&	$2 \times 10^{-5}$	&	$2 \times 10^{-6}$	\\
	Mechanical parameter  $\lambda/\kappa_{\mathrm{th}}$
		&	$1 \times 10^{3}$	&	-	&	$7 \times 10^{-10}$	&	$1 \times 10^{1}$	&	$3 \times 10^{-9}$	&	$2 \times 10^{1}$	&	$3 \times 10^{-2}$	&	$7 \times 10^{-4}$	\\
	Qubit parameter  $T_2^* \lambda/2\pi$ 
		&	$9$	&	$2$	&	$3 \times 10^{-3}$	&	$2 \times 10^{1}$	&	$4 \times 10^{-6}$	&	$1 \times 10^{-1}$	&	$6 \times 10^{-6}$	&	$3 \times 10^{-6}$	\\
	\hline
	\hline
	\end{tabular}
\caption{Parameters for a selection of implemented or proposed qubit-\resonator coupled systems for carrying out the experiment in the main text. Systems are: (1) This device; (2) Proposed levitated NV center in a magnetic field gradient (3) Implemented magnetic cantilever coupled to an NV center; (4) Proposed optimisation of device in (3); (5) Implemented NV center strain-coupled to diamond cantilever; (6) Proposed combination of the Al beam from~\cite{Teufel2009} with the qubit of~\cite{Braumuller2016}, in a field of 10~mT; (7) Proposed combination of the SiN membrane from~\cite{Yuan2015} with the qubit of~\cite{Braumuller2016}, in a field of 10~mT. The membrane is taken as supporting one arm of the qubit with length $l=300~\mu$m. (8) Proposed combination of the optimized cantilever from~\cite{Kolkowitz2012} with the qubit of~\cite{Braumuller2016}. The field from the cantilever is taken as coupling to an enclosed area of $0.1~\mu\mathrm{m}^{-2}$. None of these systems enters the bare ultrastrong coupling regime where $\lambda/\omegam>1$. However, it is possible to enter the toggled ultrastrong coupling regime where $\lambda$ exceeds both the qubit and mechanical dephasing rates.}
\label{tab:comparison}
\end{table*}

To assess the experimental feasibility of our scheme, Table~\ref{tab:comparison} presents parameters of the \resonator, the qubit, and the coupling strength for various devices that could be used to implement it. The challenge is to achieve ultrastrong coupling between qubit and \resonator without introducing either rapid dephasing of the qubit or thermal decoherence of the \resonator. Assuming a toggled coupling, this requires that the coupling constant $\lambda/2\pi$ exceeds both the qubit dephasing rate $1/T_2^*$ and the \resonator thermal dephasing rate $\kappa_{\mathrm{th}}$, as tabulated in the last two rows of the table. No existing device achieves this, although an optimized magnetic cantilever coupled to an NV center in diamond would be promising. Thus the device of Fig.~1 is attractive for investigating mesoscopic quantum interference in nanomechanics. 

\vfill

\bibliography{Bibliography,Displacemon_v23 EALNotes}

\end{document}